# Dynamics-dependent symmetries in Newtonian mechanics

**Peter Holland**
Green Templeton College
University of Oxford
Oxford OX2 6HG
England

Email: peter.holland@gtc.ox.ac.uk

**Abstract**: We exhibit two symmetries of one-dimensional Newtonian mechanics whereby a solution is built from the history of another solution via a generally nonlinear and complex potential-dependent transformation of the time. One symmetry intertwines the square roots of the kinetic and potential energies and connects solutions of the same dynamical problem (the potential is an invariant function). The other symmetry connects solutions of different dynamical problems (the potential is a scalar function). The existence of corresponding conserved quantities is examined using Noether's theorem and it is shown that the invariant-potential symmetry is correlated with energy conservation. In the Hamilton-Jacobi picture the invariant-potential transformation provides an example of a 'field-dependent' symmetry in point mechanics. It is shown that this transformation is not a symmetry of the Schrödinger equation.

## 1. Introduction

It is hard to conceive that there could be a substantive property left to discover about the classical mechanics of one-dimensional systems. Yet no anticipations of the observations presented here – symmetries of Newton's law - have been found. They derive from the following query regarding the significance of the minus sign in the classic expression $L = \frac{1}{2}m\dot{q}^2 - V(q)$ for the Lagrangian of an elementary mechanical system. Writing $L = \left(\sqrt{m/2}\dot{q}\right)^2 - \left(\sqrt{V}\right)^2$, and drawing an analogy between this difference of squares and the relativistic line element, the Lagrangian will be left invariant by a Lorentz-like transformation that mixes the functions $\sqrt{m/2}\dot{q}$ and $\sqrt{V}$. Are there associated transformations of the independent variable $t$ and dependent variable $q(t)$ that imply a symmetry of the Euler-Lagrange equations, i.e., Newton's second law? The answer is 'no' but the investigation is not fruitless: there is a symmetry of Newton's law that intertwines the square roots of the kinetic and potential energies, and it shares with relativity the property that the transformed time coordinate depends on the position. The transformation departs considerably from a relativistic symmetry in its detail, however. First, it comprises a Euclidean rotation of the 2-vector $\left(\sqrt{V}, \sqrt{m/2}\dot{q}\right)$, with the energy $E = \frac{1}{2}m\dot{q}^2 + V$ as associated invariant rather than the Lagrangian. Next, the dependence of the transformed time on the position is determined by the dynamics, via the potential. Further, the method of construction of a new solution to the dynamical equation implied by the symmetry involves an integration over the history of a known solution rather than the latter's instantaneous value. And finally, the symmetry's arena is the space of complex spacetime coordinates.

The symmetry may be decomposed into two groups: a rotation of the 2-vector that connects solutions of the same dynamical problem, and a functional variation that connects solutions corresponding to different potentials. Some of the formal aspects of these transformations are explored here, and illustrated with examples. The existence of corresponding conserved quantities is examined by applying Noether's theorem. In particular, it is shown that the rotation symmetry is correlated with the conservation of energy. Using the Hamilton-Jacobi formulation, it is shown that the rotation symmetry provides an example of a field-dependent transformation in classical point mechanics. This enables us to show that this transformation is not a symmetry of the Schrödinger equation.

## 2. The symmetries

Consider a particle of mass $m$ moving in one dimension with coordinate $q(t)$, velocity $\dot{q} = dq/dt$, initial conditions $q_0$ and $\dot{q}_0$, and potential energy $V(q)$. It is subject to Newton's law:

$$m\ddot{q} = -dV(q(t))/dq. \qquad (2.1)$$

Consider the following transformation of the independent variable $t$ and dependent variable $q$, discovered by inspection:

$$t'(t) = \int_0^t \left. \frac{d\sqrt{V(q)}/dq}{d\sqrt{V'(q')}/dq'} \right|_{q'=q(t)} dt \qquad (2.2)$$

$$\sqrt{V'(q'(t'))} = \gamma\left(\sqrt{V(q(t))} - \lambda\sqrt{m/2}\,\dot{q}\right), \qquad (2.3)$$

where $\lambda$ is a dimensionless constant parameter, $\gamma = (1+\lambda^2)^{-\frac{1}{2}}$, and the integrand in (2.2) is assumed to be an integrable function with its denominator being evaluated from (2.3). The lower limit in the integral may be fixed arbitrarily and has been chosen so that $t' = 0$ when $t = 0$ in order to easily connect the two sets of initial conditions (an alternative choice corresponds to a constant time translation which is also a symmetry of (2.1)). To show that these substitutions leave (2.1) invariant, it is convenient to first obtain from them the transformation law of the velocity. Writing $\dot{q}' = dq'/dt'$, differentiating (2.3) with respect to $t$, chain differentiating, and using (2.1) and (2.2), we get

$$\dot{q}' = \gamma\left(\dot{q} + \lambda\sqrt{2V(q)/m}\right). \qquad (2.4)$$

Together, (2.3) and (2.4) define the promised linear transformation mixing the velocity and the square root of the potential. Differentiating (2.4) with respect to $t'$ and employing (2.2) and (2.3) we deduce (2.1) written in the primed coordinates, i.e., Newton's law is covariant with respect to this transformation.

In order to obtain a formula for $q'(q(t),\dot{q}(t))$ we must specify the function $V'$. There are two possibilities: either (a) strict covariance is maintained so that $V'(q')$ is the same function of $q'$ as $V(q)$ is of $q$, i.e., $V$ is an 'invariant function': $V'(q) = V(q)$, or (b) $V$ is not an invariant function. A class (b) transformation can be decomposed into a class (a) transformation followed by the transformation

$$t'(t) = \int_0^t \left.\frac{dV(q)/dq}{dV'(q')/dq'}\right|_{q=q(t)} dt \qquad (2.5)$$

$$V'(q') = V(q) \qquad (2.6)$$

for which $\dot{q}' = \dot{q}$. For the transformation (2.5) and (2.6), denoted class (c), the function $V$ is a scalar function but not, in general, an invariant one, the converse of the situation that obtains for class (a) transformations.

In each of the cases (a) and (c), the transformation rule for $q$ may be inferred from (2.3) by inverting the function $V'(q')$ to obtain $q' = q'(V')$, which is, in general, a multivalued (and complex) function. Thus, we generally obtain a spectrum of transformation laws for $t$ and $q$, their cardinality being fixed by the number of roots $q'$. The multivaluedness generally persists when $\lambda = 0$ (class (a)) or when there is no functional variation (class (c)) and in each case we choose the root continuously connected to the identity ($t' = t$ and $q' = q$). The group property of the transformations is easily demonstrated using their infinitesimal versions described in Sec. 5. Note that a class (a) transformation is not generally a rotation in the $(q,\dot{q})$-plane. Clearly, class (a) transformations map solutions into solutions of the same dynamical problem while those of class (c) connect solutions of different dynamical problems. The mappings generally relate different regions of the complex plane and, in particular, generate complex solutions from real ones.

A new solution $q'(t')$ is obtained from a known one by a single integration over the latter's history, $q(t)$. The generally nonlinear time transformation has a geometrical interpretation: the integrand represents the fractional change in area in the $(q,\dot{q})$-plane:

$$dt'/dt = \det\left[\partial(q',\dot{q}')/\partial(q,\dot{q})\right]. \qquad (2.7)$$

These features will now be illustrated through some examples.

**3. New solutions from old for a given potential**

To illustrate the generally nonlinear and complex character of the (class (a)) transformation of $t$ and $q$ that links solutions for the same potential, we use as example the linear potential: $V(q) = mgq$ and $V'(q') = mgq'$. Then $q(t) = q_0 + \dot{q}_0 t - \frac{1}{2}gt^2$ and, from (2.3),

$$q' = \gamma^2\left(\sqrt{q} - \lambda\sqrt{1/2g}\,\dot{q}\right)^2. \qquad (3.1)$$

We aim to construct the latter as a function of the new time coordinate (2.2):

$$t'(t) == \int_0^t \sqrt{q'/q}\, dt = \gamma\left\{t - \lambda\sqrt{2/g}\left[\sqrt{q_0 + \dot{q}_0 t - \tfrac{1}{2}gt^2} - \sqrt{q_0}\right]\right\}. \qquad (3.2)$$

Inverting this formula to obtain $t(t')$ and substituting in (3.1) we indeed obtain the general solution for the linear potential in the primed coordinates with initial conditions connected by (2.3) and (2.4). From (3.2) with $\lambda \neq 0$ and real $t$ and $q$, $t'$ and hence $q'$ are real only for an epoch $(0,t)$ in which $q \geq 0$.

## 4. Conversion between dynamical problems

Here we give two simple examples illustrating how a class (c) transformation of $t$ and $q$ provides a method to solve a dynamical problem using a known solution pertaining to a different potential function.

*Harmonic oscillation from uniform acceleration*

Suppose $V(q) = mgq$ and $V'(q') = \tfrac{1}{2}m\omega^2 q'^2$. Starting from the uniformly accelerating trajectory $q(t) = q_0 + \dot{q}_0 t - \tfrac{1}{2}gt^2$ we wish to deduce the trajectory of the harmonic oscillator implied by (2.6), namely $q'(t') = \omega^{-1}\sqrt{2gq(t(t'))}$. For the transformed time coordinate we have

$$t'(t) = \frac{\sqrt{g/2}}{\omega}\int_0^t q(t)^{-1/2}\, dt = \frac{1}{i\omega}\log\left(\frac{i\sqrt{2gq(t)} + \dot{q}(t)}{i\sqrt{2gq_0} + \dot{q}_0}\right). \qquad (4.1)$$

Then, solving (4.1) to obtain $q(t)$ in terms of $t'$, we readily find the harmonic oscillator orbit

$$q'(t') = q'_0 \cos\omega t' + (\dot{q}'_0/\omega)\sin\omega t' \qquad (4.2)$$

where $q'_0 = \sqrt{2gq_0}/\omega$, $\dot{q}'_0 = \dot{q}_0$. Using the formula

$$E = \left(i\sqrt{2gq} + \dot{q}\right)\left(-i\sqrt{2gq} + \dot{q}\right) = \left(i\sqrt{2gq_0} + \dot{q}_0\right)\left(-i\sqrt{2gq_0} + \dot{q}_0\right) \qquad (4.3)$$

it is easy to see from (4.1) that $t'$ and $q'$ become complex when $q < 0$.

*Displaced harmonic oscillation from harmonic oscillation*

In this example a solution is constructed from one depending on a different (lesser) number of parameters. Let $V(q) = \tfrac{1}{2}m\omega^2 q^2$ and $V'(q') = mgq' + \tfrac{1}{2}m\omega'^2 q'^2$. Writing the oscillating trajectory as $q = A\cos(\omega t + B)$, we have

$$t'(t) = \omega^2 \int_0^t q\left(g^2 + \omega^2\omega'^2 q^2\right)^{-1/2} dt = \frac{1}{\omega'}\sin^{-1}\left(\frac{A\omega\omega'\sin(\omega t + B)}{\sqrt{g^2 + A^2\omega^2\omega'^2}}\right) - \frac{B'}{\omega'} \qquad (4.4)$$

where $\sin B' = A\omega\omega' \sin B \left(g^2 + A^2\omega^2\omega'^2\right)^{-1/2}$. Then,

$$q' = -g\omega'^{-2} + \omega'^{-2}\sqrt{g^2 + A^2\omega^2\omega'^2}\cos(\omega't' + B'). \qquad (4.5)$$

This gives the functional dependence of the displaced harmonic oscillator with all the parameters of the two dynamical problems contributing to the transformed initial condition $q'_0$.

## 5. Noether charges

Noether's (first) theorem concerns continuous transformations $t,q \to t',q'$ of a dynamical system that leave the action $\int L dt$ quasi-invariant:

$$\int L(q',\dot{q}',t')dt' = \int L(q,\dot{q},t)dt + \int (d\Gamma/dt)dt \qquad (5.1)$$

for some function $\Gamma$. For each transformation, the old and new actions generate the same Euler-Lagrange equation upon applying Hamilton's principle. The theorem establishes a correlation between these 'variational symmetries' and certain constants of the motion, denoted 'Noether charges'. To examine whether this framework may be applied to the symmetries we have presented we require a version of the theorem sufficiently general to encompass velocity-dependent transformations (for details see [1,2]) and we need to determine whether the symmetries are indeed variational ones.

We consider first a class (a) symmetry. For infinitesimal $\lambda$, we write the transformations of the independent and dependent variables as follows:

$$t' = t + \lambda\xi(q,\dot{q},t), \quad q'(t') = q(t) + \lambda\eta(q,\dot{q},t). \qquad (5.2)$$

The induced infinitesimal transformation of the velocity is

$$\dot{q}' = \dot{q} + \lambda\left(\frac{d\eta}{dt} - \frac{d\xi}{dt}\dot{q}\right) \qquad (5.3)$$

where

$$\frac{d}{dt} = \frac{\partial}{\partial t} + \dot{q}\frac{\partial}{\partial q} + \ddot{q}\frac{\partial}{\partial \dot{q}}. \qquad (5.4)$$

For a transformation under which $L$ is an invariant function, such as a class (a) symmetry, the quasi-invariance of the action under the transformation (5.2) is ensured by the local condition

$$L(q',dq'/dt',t')\left(1 + \lambda\frac{d\xi}{dt}\right) = L(q,dq/dt,t) + \lambda\frac{d\Lambda}{dt} \qquad (5.5)$$

where the function $\Lambda$ depends on $t,q,\dot{q}$. Expanding the left-hand side of (5.5) to first order in $\lambda$, and subjecting the function $q$ to the Euler-Lagrange equation (2.1), we obtain

$$\xi \frac{\partial L}{\partial t} + \eta \frac{\partial L}{\partial q} + \left( \frac{d\eta}{dt} - \frac{d\xi}{dt} \dot{q} \right) \frac{\partial L}{\partial \dot{q}} + L \frac{d\xi}{dt} = \frac{d\Lambda}{dt}. \qquad (5.6)$$

The variational symmetries are given by the solutions $\xi, \eta$ to this equation. Noether's theorem comprises the observation that (5.6) may be reformulated as a conservation law, $dP/dt = 0$, where the Noether charge is given by [1,2]

$$P(q, \dot{q}, t) = L\xi + \frac{\partial L}{\partial \dot{q}}(\eta - \dot{q}\xi) - \Lambda \qquad (5.7)$$

for each variational symmetry $\xi, \eta$. Thus, (5.6) provides both a means to discover a class of symmetries and a formula for the associated set of conserved quantities.

In our case we know the transformations of $t$ and $q$ and aim to determine whether their infinitesimal forms obey (5.6) for a suitably defined function $\Lambda$. To establish these forms, we compare the infinitesimal transformation of the potential, which is, for a general invariant function,

$$V'(q') = V(q) + \lambda \frac{\partial V}{\partial q} \eta(q, \dot{q}, t), \qquad (5.8)$$

with the first order version of the specific transformation (2.3). We obtain

$$\eta(q, \dot{q}) = \dot{q} f(q), \quad f = -\sqrt{m/2} \left( \partial \sqrt{V} / \partial q \right)^{-1}. \qquad (5.9)$$

From the formula (2.7) we have

$$\frac{d\xi}{dt} = \frac{\partial \eta}{\partial q} \qquad (5.10)$$

whence $\xi(q) = f(q) - f(q_0)$. Hence, using (5.9), the infinitesimal form (5.2) of the class (a) transformation is

$$\left. \begin{array}{l} t' = t - \lambda \sqrt{m/2} \left[ \left( \partial \sqrt{V(q)} / \partial q \right)^{-1} - \left( \partial \sqrt{V(q_0)} / \partial q_0 \right)^{-1} \right] \\ q'(t') = q(t) - \lambda \sqrt{m/2} \dot{q} \left( \partial \sqrt{V} / \partial q \right)^{-1}. \end{array} \right\} \qquad (5.11)$$

Substituting these expressions for $\xi$ and $\eta$, it is easy to confirm that (5.6) is indeed obeyed, with $\Lambda = Lf$. A class (a) transformation is therefore a variational symmetry.

The corresponding conserved charge (5.7) is $P = Ef_0$, i.e., the energy. This example supplies an instructive antidote to the view, implicit in many presentations, that the conservation of energy is uniquely correlated with invariance with respect to time translations. That correlation certainly exists but the association is not unique, even with respect to the same Lagrangian.

To examine whether Noether's theorem can be applied to a class (c) symmetry we must add the infinitesimal functional variation $\delta V(q) = V'(q) - V(q)$ to the left-hand side of (5.5) and to the right-hand side of (5.8). Comparing the latter with the first order limit of (2.6), and using (2.7), we obtain the equations

$$\delta V(q) + \lambda \eta \frac{\partial V}{\partial q} = 0, \quad \frac{d\xi}{dt} = \frac{\partial \eta}{\partial q}, \tag{5.12}$$

which connect $\eta$, $\xi$ and $\delta V$. The condition for a quasi-invariant action reduces to

$$L \frac{\partial \eta}{\partial q} = \frac{d\Lambda}{dt}. \tag{5.13}$$

To obtain a conserved quantity, $\delta V$ must be a total time derivative, $dF(q,\dot{q},t)/dt$, which generally it is not. An exception occurs for a variation for which $\eta = $ constant so that $F = m\dot{q}\eta$, $\Lambda = 0$, and $\xi = $ constant. This corresponds to a rigid infinitesimal shift of the profile of the potential (so, for the harmonic oscillator, one has $V'(q) = \frac{1}{2} m\omega^2 (q - \lambda\eta)^2$). The corresponding conserved charge is once again the energy, $P = -E\xi$. This is expected since the symmetry includes a constant time translation with respect to which the potential is invariant (but the associated uniform change in $q$ is not accompanied by momentum conservation). In general, however, a class (c) transformation is not a variational symmetry; it provides an instance of a dynamical symmetry (a symmetry of the Euler-Lagrange equation) to which Noether's theorem does not apply. For the harmonic oscillator, an example of a non-variational infinitesimal functional change would be a shift in the frequency: $V'(q) = \frac{1}{2} m(\omega + \lambda\Omega)^2 q^2$.

**6. Field-dependent transformation**

To examine whether the classical symmetry (a) is also a symmetry of the Schrödinger equation, we first translate our results into the field-theoretic Hamilton-Jacobi language where $t$ and $q$ (now denoted $x$) are independent variables, and dynamical evolution is described by the Hamilton-Jacobi equation

$$\frac{\partial S}{\partial t} + \frac{1}{2m}\left(\frac{\partial S}{\partial x}\right)^2 + V(x) = 0. \tag{6.1}$$

Following (5.11), the invariance group of (6.1) corresponding to a class (a) symmetry is, to first order,

$$\left.\begin{array}{l} t' = t + \lambda f(x), \quad x' = x + \lambda f(x) \dfrac{1}{m}\dfrac{\partial S}{\partial x}, \\[6pt] V'(x') = V(x) - \lambda \sqrt{\dfrac{2V}{m}} \dfrac{\partial S}{\partial x}, \quad S'(x',t') = S(x,t) + \lambda f(x) L(x,t), \end{array}\right\} \tag{6.2}$$

where $f(x) = -\sqrt{m/2}\left(d\sqrt{V}/dx\right)^{-1}$ and $L(x,t) = (1/2m)(\partial S/\partial x)^2 - V$. The transformation of $S$ is in accord with the quasi-invariance of the action found in Sec. 5 (the latter being the solution of (6.1) whose dependent non-additive constant is the initial position). The functional dependence of $x'$ renders this transformation 'field-dependent', that is, the coordinate change involves the field ($S$) that is subject to that change. Field-dependent symmetries were discovered in the context of fluid-mechanical models involving, in particular, Hamilton-Jacobi-like equations describing fluids possessing certain types of constitutive relations [3]. Following that lead, we might append to the field-theoretic description of the classical-mechanical system the continuity equation

$$\frac{\partial \rho}{\partial t} + \frac{\partial}{\partial x}\left(\rho \frac{1}{m}\frac{\partial S}{\partial x}\right) = 0. \tag{6.3}$$

To first order, this equation is covariant under (6.2) when the density transforms as

$$\rho'(x',t') = \rho(x,t) - \lambda(1/m)\rho f \partial^2 S/\partial x^2. \tag{6.4}$$

Note that $t'$, $x'$, $S'$ and $\rho'$ are either all real or all complex. This theory lies outside the class of fluid theories just mentioned, which may be characterized as the case where the dependence of $V$ on $x$ derives solely from a dependence on $\rho$. Clearly, the transformation rules for $V$ and $\rho$ in our case preclude such a functional relation.

Using the formula (6.4), and considering the quantal continuity equation, it is easy to see that the transformation (6.2) is not a symmetry of the Schrödinger equation. In the path integral representation of the wavefunction, the propagator is built from a superposition of elementary amplitudes, each determined by the classical action. Adding the infinitesimal total derivative $\lambda\, d\Lambda/dt$ to the Lagrangian multiplies each amplitude by the same factor, $1 + \lambda(i/\hbar)(\Lambda - \Lambda_0)$. This in turn induces an infinitesimal gauge transformation of the wavefunction:

$$\psi'(x',t') = \psi(x,t)\left(1 + \lambda\frac{i}{\hbar}Lf\right) \tag{6.5}$$

where $x'$ and $t'$ are given by (6.2). In a domain where $f$ is real, the quantal phase therefore transforms like the classical Hamilton-Jacobi function $S$ and the density $|\psi|^2$ is invariant. But the latter property contradicts the transformation (6.4) of a density necessary to maintain covariance of the continuity equation for such a phase change. How the transformation might be modified to generate a Schrödinger symmetry is an open question.

## 7. Conclusion

We have exhibited apparently unremarked symmetries of one-dimensional classical dynamics that share a common dynamics-dependent nonlinear time transformation in the complex plane. A solution is built by a single integration over the history of a known solution pertaining to the same or a different dynamical problem. In the former case (a) the construction involves a technique reminiscent of that employed to factor the Hamiltonian in supersymmetric quantum mechanics but it has no obvious connection with that method (or with supersymmetric classical mechanics [4]). Potential-dependent transformations of this sort also seem to lie outside the realm of higher-order quantum symmetries [5]. Examining the theory's computational merits, extending it to higher dimensions, and understanding its provenance within quantum mechanics, are subjects for future enquiries.